\documentstyle [11pt] {article}

\begin{document}
\title{On the analogy between electromagnetism and turbulent hydrodynamics}
\author{Haralambos Marmanis\\
Department of Theoretical and Applied Mechanics\\
University of Illinois at Urbana-Champaign\\
Urbana, Illinois 61801-2935, USA.}
\date{\today}
\maketitle

\begin{abstract}
In this note, we propose an exegesis of the Maxwell equations for
electromagnetism. We begin with an analogy between the homogeneous Maxwell
equations and the equations needed to describe the vorticity field of an
incompressible inviscid fluid. We suggest that the inhomogeneous equations are
analogous to two equations valid in turbulent hydrodynamics. Once the analogy
is completed we give the mechanical analogue of the Poynting vector and we
explain the influence of a long solenoid on the motion of a charged particle.
\end{abstract}

\section{Introduction}
The nature of the electromagnetic field has been a puzzle since the foundations
of the subject by Maxwell and even earlier (Whittaker, 1951). Scientists have
tried to imagine the electric and magnetic field in terms of other physical
circumstances where insight is provided by the phenomenon itself. In order to
do this people had postulated the subsistence of a very fine medium, the
so-called `aether'. This seemed to be a legitimate act, because the
electromagnetic theory based on Maxwell's equations predicts the existence of
waves. However, no experimental proof  that the aether really exists could be
found; the negative result of the Michelson-Morley experiment has been verified
many times with various modifications. Thus the concept of aether has been
abandoned. Nevertheless the problem of what is the nature of the
electromagnetic field cannot be said to have been dealt with completely. What
has happened is that physicists simply assumed the existence of the fields and
postponed the resolution of the problem. Feynman (1964) devotes a whole section
 of his lectures on this matter.

In the sequel we propose an analogy between the vector fields of
electromagnetism and the hydrodynamical fields of a  turbulent fluid flow. In
this analogy the field quantities are envisaged as the fundamental entities
whereas the charges and currents are byproducts of the former. In that sense
this work is a reminiscence of Faraday's theory were electric charges were to
be regarded as epiphenomena having no independent or substantial existence. The
electric current, accordingly, was to be viewed not as manifesting the flow of
actually existing electrical fluids but rather as constituting an ``axis of
power'', reflecting the dynamics of ``a certain condition and relation of
electrical forces.'' (Siegel, 1991).

Einstein suggested that special kinds of non-linear fields might exist, having
modes of motion in which there would be pulse-like concentrations of fields,
which would stick together stably, and would act almost like small moving
bodies. Heisenberg had a similar point of view. Misner (1956) proposed that
electromagnetism was a property of  curved empty space. Rainich (1925) already
long before had shown under what conditions a curvature of space-time can be
regarded as due to electromagnetic field. Finally Misner and Wheeler (1957)
presented gravitation, electromagnetism, unquantized charge, and mass as
properties of curved empty space. Furthermore, the particle theory developed by
M.Vigier and his collaborators (Hillion, Halbwachs, Lochak), is also along the
same line of thought but is focused primarily on the quantum mechanical
phenomena. It is the purpose of this paper to propound that the space-time
manifold can be represented as the manifold of a turbulent fluid flow. The
latter is possible to initiate profound changes about the interpretation of
 electromagnetism and quantum mechanics.

The analogical approach presented here, it will provide us with a mechanical
representation, which will be helpfull in the task of understanding the nature
of electromagnetism by using more familiar terms. In section 2, we present the
Maxwell equations and the equations of hydrodynamics for an inviscid
incompressible fluid, in terms of the velocity and the Bernoulli energy
function, and in terms of the vorticity and the Lamb vector. The resemblance
between these equations will be the starting point of our analogy. In section 3
a correspondence between electromagnetic and hydrodynamical quantities is
established, where the vector and scalar potential of the electromagnetic field
appear to be as `real' as the magnetic and the electric field. We argue that
all the electromagnetic variables can be interpreted as hydrodynamical
variables of a turbulent flow field. In section 4, we discuss the origin of the
Poynting vector and the effect of a long solenoid on the motion of a charged
particle passing nearby, according to the mechanical analogue.

\section{The equations of Maxwell and of hydrodynamics}

The Maxwell equations have different coefficients according to the system of
units that is chosen. If  we choose the electrostatic (esu) system, the Maxwell
equations for sources in vacuum can be written as follows (Jackson, 1975)
\begin{eqnarray}
\nabla \cdot {\bf B} & = & 0\, , \\
\frac{\partial\, {\bf B}}{\partial t} & = & -\, \nabla \times {\bf E}\, , \\
\nabla \cdot {\bf E} & = & 4 \pi\, \rho\, , \\
\frac{\partial\, {\bf E}}{\partial t} & = & c^{2}\, \nabla \times {\bf B}\, -\,
4 \pi\, {\bf J}\, ,
\end{eqnarray}
where ${\bf B}({\bf x}, t)$ is the magnetic field, ${\bf E}({\bf x}, t)$ is the
electric field, $\rho ({\bf x}, t)$ is the charge density, and ${\bf J}({\bf
x}, t)$ is the current. Moreover, the fields ${\bf B}$ and ${\bf E}$ form a
six-component system, but not all of these components are entirely independent;
this is implicitly expressed in equations (1) and (2). In other words, we can
find a more economical description of the fields with fewer components. This is
done by introducing the vector potential ${\bf A}({\bf x}, t)$ and the scalar
potantial $\phi({\bf x}, t)$. It is easy to see that the substitutions
\begin {equation}
{\bf B}\, =\, \nabla \wedge {\bf A}\, ,
\end{equation}
\begin{equation}
{\bf E}\, =\, - \frac{\partial\, {\bf A}}{\partial t}\, - \nabla \phi
\end{equation}
yield (1) and (2) as identities.

On the other hand, in hydrodynamics the flow of an incompressible inviscid
fluid, of constant density $\rho$, is governed by the Euler equations. The
latter can be written in the following form
\begin{equation}
\frac{\partial\, {\bf u}}{\partial t}\, =\, -\, ({\bf w} \wedge {\bf u}) -\,
\nabla \left( \frac{p}{\rho} + \frac{u^{2}}{2} \right)\,  ,
\end{equation}
where ${\bf u}({\bf x}, t)$ is the velocity field, ${\bf w}({\bf x}, t)$ is the
vorticity field, and $p({\bf x}, t)$ is the pressure field. The vector product
of the vorticity with the velocity is called the Lamb vector, and it will be
denoted as
\[ {\bf l}({\bf x}, t)\, =\, {\bf w} \wedge {\bf u}\, . \]
The quantity in the parenthesis of the second term in the r.h.s. of (7) is
called the Bernoulli energy function or total `head', and will be denoted as
\[ \Phi({\bf x}, t)\, =\, \frac{p}{\rho} + \frac{u^{2}}{2}\, . \]
The Euler equations are accompanied by the continuity equation which, for an
incompressible fluid, is reduced to
\begin{equation}
\nabla \cdot {\bf u}\, =\, 0\, .
\end{equation}
The vorticity field, defined as
\begin{equation}
{\bf w}\, =\, \nabla \wedge {\bf u}\, ,
\end{equation}
obeys the following two equations
\begin{equation}
\nabla \cdot {\bf w}\, =\, 0\, ,
\end{equation}
\begin{equation}
\frac{\partial\, {\bf w}}{\partial t}\, =\, - \nabla \wedge {\bf l}\, .
\end{equation}
Equation (11) can be easily derived by taking the $curl$ of (7).

\section{The hydrodynamical analogy}
Now, by comparing equations (1) and (2) with equations (10) and (11),
respectively, we observe that are the same if ${\bf B}$ corresponds to ${\bf
w}$, and ${\bf E}$ corresponds to ${\bf l}$. In addition, by comparing
equations (5) and (6) with (9) and (7), respectively, we see that the analogy
can be extended, so that it includes the potentials as well. In particular, the
comparison suggests that the vector potential ${\bf A}$ corresponds to the
velocity field ${\bf u}$, and the scalar potential $\phi$ to the Bernoulli
energy function $\Phi$. The complete correspondence, between the
electromagnetic fields and their hydrodynamical analogues, can be summarized as
follows
\begin{center}
\begin{tabular}{|c|c|}
\hline
Electromagnetism & Hydrodynamics \\ \hline
${\bf A}({\bf x}, t)$ & ${\bf u}({\bf x}, t)$ \\ \hline
$\phi ({\bf x}, t)$ & $\Phi ({\bf x}, t)$ \\ \hline
${\bf B}({\bf x}, t)$ & ${\bf w}({\bf x}, t)$ \\ \hline
${\bf E}({\bf x}, t)$ & ${\bf l}({\bf x}, t)$ \\ \hline
\end{tabular}
\end{center}
 From the above analogy, the magnetic field appears to have a rotatory
character. This was known to Maxwell himself, since the very begining of his
investigations. He was aware of the fact that magnetism produces, at least, one
rotatory effect, i.e. the rotation of the plane of polarized light when
transmitted along the magnetic lines (Faraday rotation). Even though later he
seemed to decline his theory of molecular vortices, he never did give up the
belief that there was a real rotation going on in the magnetic field.
By applying the divergence operator on both sides of (7), we get
\begin{equation}
\nabla \cdot {\bf l}({\bf x}, t)\, =\, - \nabla^{2}\, \Phi\, .
\end{equation}
One can also express the divergence of the Lamb vector in the form
\begin{eqnarray}
\nabla \cdot {\bf l}({\bf x}, t) & = & {\bf w} \cdot {\bf w}\, +\, {\bf u}
\cdot \nabla^{2} {\bf u} \\
   & = & {\bf w} \cdot {\bf w}\, -\, {\bf u} \cdot \nabla \wedge {\bf w}\; .
\end{eqnarray}
In general, the Laplacian of $\Phi$ will be a function of position and time, so
let us call this function the {\em turbulent charge density} and denote it as
$n({\bf x}, t)$, then we can write
\begin{equation}
\nabla \cdot {\bf l}({\bf x}, t)\, =\, - \nabla^{2}\, \Phi\, \equiv\, 4 \pi\,
n({\bf x}, t)\, ,
\end{equation}
where the $4 \pi$ proportionality factor is introduced for later convenience.
It is clear that the function $n({\bf x}, t)$ will be significantly greater in
a turbulent flow than in a laminar flow. For,  in a turbulent flow, the
enstrophy is larger, mainly due to the stretching of vortex filaments, the
velocity is larger and the flexion vector is also larger. Thus the designation
of $n({\bf x}, t)$ as {\em turbulent} is justified. On the other hand equation
(15) reminds the Poisson equation of electromagnetism, which connects the
electric potential $\phi ({\bf x}, t)$ and the electric charge density $\rho
({\bf x}, t)$ as follows
\begin{equation}
\nabla^{2} \phi\, =\, -\, 4 \pi\, \rho ({\bf x}, t)\, .
\end{equation}
Because of this resemblance we envision $n({\bf x}, t)$ as a {\em charge}
density.

Let us note that there are two points of view about the interpretation of the
above equation which disclose two points of view about the nature of physics,
in general. According to the first interpretation the space-time continuum is
not inherently connected with the existence of the fields and the particles.
These entities have a right of existence on their own. The description of
physical phenomena requires the space-time arena, the fields and the particles.
The second interpretation according to Misner \& Wheeler (1957) can be
described as follows: ``There is nothing in the world except empty curved
space. Matter, charge, electromagnetism, and other fields are only
manifestations of the bending of space. Physics is geometry''. Herein we adopt
the later point of view, thus in equation (16) the electric potential is the
cause and the electric charge the effect. This is plain for equation (14) where
we introduced the notion of turbulent charge density as a result of the
curvature of the Bernoulli energy function.

At this point, one could say that the nature of the electromagnetic field is
that of an incompressible inviscid  turbulent fluid flow, which in addition to
the known equations described above, obeys also the following equation
\begin{equation}
\frac{\partial\, {\bf l}}{\partial t}\, =\, c^{2}\, \nabla \wedge {\bf w}\, -\,
4 \pi\, {\bf I}\; ,
\end{equation}
where ${\bf I}({\bf x}, t)$ is given as a function of $n({\bf x}, t)$, through
the relation
\begin{equation}
\frac{\partial\, n}{\partial t}\, +\, \nabla \cdot {\bf I}\, =\, 0\; .
\end{equation}
J. J. Thomson (1931) showed that in the case of a homogeneous and isotropic
flow with a large number of vortex filaments, the Lamb vector obeys equation
(17) with ${\bf I}= 0$; as required by homogeneity. The idea of delineating
turbulent flows as a large irrotational region which occupies most of the flow
field and narrow regions of concentrated vorticity came almost two decades
later (Onsager, 1949; Kida, 1975). Numerical evidence supporting the above
picture, especially for the small scales, has been presented recently (Siggia,
1981; Kerr, 1985; Vincent \& Meneguzzi, 1991; Jim$\acute{e}$nez {\em et al.},
1993; Kida, 1993).

\section{Two old questions}
One special characteristic, that didtinguishes our analogy from previous
attempts, is the mechanical exegesis of a few important questions in
electromagnetism. We will illustrate this by two concrete examples. The first
case is the derivation of the Poynting vector
\begin{equation}
{\bf S}\; =\; c^{2}\; {\bf E} \wedge {\bf B}\, .
\end{equation}
 The latter gives the rate at which the field energy moves around in space.

In general, the conservation of energy for electromagnetism is written as
\begin{equation}
\frac{\partial U}{\partial t}\, +\, \nabla \cdot {\bf S}\, =\, {\bf E} \cdot
{\bf J}\, ,
\end{equation}
where $U$ is given by
\begin{equation}
U\, =\, \frac{1}{2}\, {\bf E} \cdot {\bf E}\, +\, \frac{c^{2}}{2}\, {\bf B}
\cdot {\bf B}\, .
\end{equation}
The above equation corresponds to the evolution equation of the enstrophy in a
fluid containing a large number of vortex filaments. Let us examine the case of
fields in vacuum with no charges, i.e. the homogeneous case considered by
Thomson, then $\rho=0$ and ${\bf J}\, =\, 0$ in (3) and (4) respectively.
By multiplying equation (11) with ${\bf w}$, we get
\begin{equation}
\frac{\partial}{\partial t}\, \frac{{\bf w}^{2}}{2}\, =\, -\, {\bf w} \cdot
\nabla \wedge {\bf l}\, .
\end{equation}
However
\begin{equation}
 -\, {\bf w} \cdot \nabla \wedge {\bf l}\, =\, -\, \nabla \cdot ({\bf l} \wedge
{\bf w})\, +\, {\bf l} \cdot \nabla \wedge {\bf w}\, ,
\end{equation}
therefore (23) can be written as
\begin{equation}
\frac{\partial}{\partial t}\, \frac{{\bf w}^{2}}{2}\, +\, \nabla \cdot ({\bf l}
\wedge {\bf w})\, =\, {\bf l} \cdot \nabla \wedge {\bf w}\, .
\end{equation}
The term on the r.h.s. has been called the turbulence creating term and its
significance to turbulence theory has been investigated  by Theodorsen (1952).
Substituting the equation (19) derived by Thomson (1931) and rederived by
Marmanis (1996), we have
\begin{equation}
\frac{\partial}{\partial t}\, \frac{{\bf w}^{2}}{2}\, +\, \frac{1}{c^{2}}\,
\frac{\partial}{\partial t}\, \frac{{\bf l}^{2}}{2}\, +\, \nabla \cdot ({\bf l}
\wedge {\bf w})\, =\, 0
\end{equation}
or
\begin{equation}
\frac{\partial}{\partial t}\, \left( \frac{c^{2}\, {\bf w}^{2}}{2}\, +\,
\frac{{\bf l}^{2}}{2} \right)\, +\, c^{2}\, \nabla \cdot ({\bf l} \wedge {\bf
w})\, =\, 0\, ,
\end{equation}
which according to our analogy is exactly (21) if the Poynting vector
corresponds to
\begin{equation}
{\bf S}\; =\; c^{2}\; {\bf l} \wedge {\bf w}\, .
\end{equation}
Notice that if this is true the Poynting vector is aligned with the vector
potential. It is also interesting that the electrostatic energy has as
mechanical analogue a special combination of the three most important
quantities in any turbulent flow, namely the kinetic energy, the enstrophy and
the magnitude of the helicity density. In particular, we have
\begin{equation}
{\bf l}^{2}\, =\, {\bf w}^{2}\, {\bf u}^{2}\, -\, ({\bf u} \cdot {\bf w})^{2}\;
{}.
\end{equation}
Furthermore the Poynting vector is given also by these important quantities as
\begin{equation}
{\bf l} \wedge {\bf w}\, =\, ({\bf w}^{2}) {\bf u}\, -\, ({\bf u} \cdot {\bf
w}) {\bf w}\; .
\end{equation}
The above equation tells us a little more about the energy flow according to
our analogy. It seems that the energy flow increases in the direction of the
vector potential when the magnetic energy increases and decreases in the
direction of the magnetic field when the electromagnetic helicity decreases. A
special case is the two-dimensional flows. In the latter case the helicity is
zero and the energy flow is given solely in terms of the enstrophy and the
velocity. However, the enstrophy in a two-dimensional inviscid flow is
conserved, therefore any change of the energy flow will be due to the velocity.

Our second example is the influence of a long solenoid on the motion of charged
particles (Feynman, 1964). Classically, i.e. not quantum mechanically, the
force depends only on ${\bf B}$. That is, in order to know that the solenoid is
carrying current, the particle must go through it. Quantum mechanics predicts
an influence on the motion given in terms of the magnetic change in phase.
According to the picture proposed herein, the influence of the field is
explained as follows. As the particle approaches the solenoid, it will be
exposed to the flow field created by it. The latter can be approximated by that
of a vortex tube. Therefore the influence will be more or less significant
according to the value which describes the strength of the vortex tube. This
quantity is just the integral of the velocity on a path surrounding the vortex
tube, that is
\begin{equation}
\oint {\bf u} \cdot d{\bf s}\; ,
\end{equation}
where $d{\bf s}$ is the differential arc element along the path. This result is
in accordance with the result of quantum mechanics, i.e.
\begin{equation}
\delta\, =\, \frac{q}{\hbar} \oint {\bf A} \cdot d{\bf s}\; ,
\end{equation}
where $\delta$ is the magnetic change in phase of the wave function attributed
to the particle,
$q$ is the charge and $h$ the Plank's constant. It is also remarkable that the
vector potential of a
long solenoid behaves in exact the same way as the velocity field of a vortex
filament, i.e. they
both decrease as $r^{-1}$ with increasing r.

\section{Discussion}

We decsribed the analogy between the equations of electromagnetism and the
equations of turbulent
hydrodynamics. There is a one-to one correspondence between quantities in the
two cases. This leads us
to interpret classical electromagnetism as a turbulent flow field. Of course,
such a statement has
consequences proportional to its generality and its vagueness. For example, it
has long been recognized
that turbulent fluid flows have an intermittent character, especially at the
small scales (Batchelor \& Townsend, 1949).
The separation of the various sources of intermittency is insufficiently
recognized in the literature (Kraichnan, 1991).
For reasons that we explain elsewhere (Marmanis, 1993), we claim that the
intermittency of turbulent fluid
flows should be accredited to intermittency effects intrinsic to the
dissipation range. Therefore it is
tempting to propose a new set of equations for electromagnetism. The modified
Maxwell equations will read
\begin{eqnarray}
\nabla \cdot {\bf B} & = & 0\, , \\
\frac{\partial\, {\bf B}}{\partial t} & = & -\, \nabla \times {\bf E}\, +\,
\hbar \nabla^{2} {\bf B} , \\
\nabla \cdot {\bf E} & = & 4 \pi\, \rho\, , \\
\frac{\partial\, {\bf E}}{\partial t} & = & c^{2}\, \nabla \times {\bf B}\, -\,
4 \pi\, {\bf J}\, +\, \hbar \nabla^{2} {\bf E} ,
\end{eqnarray}
where $\hbar$ is Plank's constant per unit mass.

In conclusion, an analogy between electromagnetism and hydrodynamics is
presented. The analogy by Thomson (1931) has been very similar to ours as far
as the construction of equation (4) is concerned. However, the interpretation
given herein is different than Thomson's interpretation. The early ideas of
Faraday, Maxwell, Rowland and others about the nature of electromagnetism are
now illuminated
more than ever; under the new perspective of turbulent hydrodynamics.

\section*{Acknowledgements}
The author wants to acknowledge the valuable communication with Dr. R.H.
Kraichnan and Prof. A. Leggett,
as well as the helpful discussions with Prof. R. Adrian, Prof. H. Aref, Prof.
G.I. Barenblatt, Dr. Meleshko, and Prof. R. Moser.


\end{document}